%% file: main.tex
\documentclass[]{fairmeta}

\usepackage{hyperref}

\usepackage{amsmath}
\usepackage{amsthm}
\usepackage{xcolor}
\usepackage{color}

\usepackage{cleveref}
\usepackage{mathtools}
\usepackage{subcaption}

\usepackage{cleveref}
\usepackage{tikz}
\usepackage{pgfplots}
\usepackage{csvsimple}

\pgfplotsset{compat=newest}
\usepgfplotslibrary{colorbrewer}
\pgfplotsset{
    cycle list/Paired,
    cycle multiindex* list={
        mark list*\nextlist
        Paired\nextlist
    },
    grid style={dotted,black},
}

\input{macros}

\title{System-2 Recommenders\\ \large 
Disentangling Utility and Engagement 
in Recommendation Systems via Temporal Point-Processes}

\author{Arpit Agarwal}
\author{Nicolas Usunier}
\author{Alessandro Lazaric}
\author{Maximilian Nickel}

\affiliation{FAIR at Meta}

\abstract{

Recommender systems are an important part of the modern human experience whose influence ranges from the food we eat to the news we read. Yet, there is still debate as to what extent online recommendation platforms are \emph{aligned} with the goals of their users. A core issue fueling this debate is the challenge of inferring a user’s utility based on their engagement signals such as likes, shares, watch time etc., which are often the primary metric used by platforms to optimize content. 
This is because users' utility-driven decision-processes (which we refer to as \emph{System-2}), e.g., reading news that are accurate and relevant for them,
are often confounded by their impulsive or unconscious decision-processes (which we refer to as \emph{System-1}), e.g., spend time on click-bait news articles. 
As a result, it is difficult to infer 
whether an observed engagement is utility-driven or impulse-driven. 
In this paper we explore a new approach to recommender systems where we infer user’s utility based on their return probability to the platform rather than engagement signals. 
This approach is based on 
the intuition that users tend to return to a platform in the long run if it creates utility for them, while pure engagement-driven interactions, i.e., interactions that do not add meaningful utility, may affect user return in the short term but will not have a lasting effect. 
For this purpose, we propose a generative model in which past content interactions impact the arrival rates of users based on a self-exciting Hawkes process. These arrival rates to the platform are a combination of both System-1 
and System-2 decision processes. The System-2 arrival intensity depends on the utility drawn 
from past content interactions and has a long lasting effect on return probability. In contrast, System-1 arrival intensity depends on the instantaneous gratification or \emph{moreishness} and tends to vanish rapidly in time.
We show analytically that given samples from this model it is provably possible to disentangle 
the System-1 and System-2 decision-processes
and thus infer user's utility, thereby allowing us to optimize content based on it. We conduct experiments on synthetic data to demonstrate the effectiveness of our approach over engagement optimization.

}

\date{\today}
\correspondence{\email{aarpit@meta.com}}

\begin{document}

\maketitle

\input{intro}

\input{problem}

\input{identifiability}

\input{expts}

\input{conclusion}

\bibliographystyle{assets/plainnat}
\bibliography{bib}

\clearpage
\newpage
\beginappendix
\input{appendix}


\end{document}

%% file: macros.tex
\newcommand{\Ex}{\mathbb{E}}
\newcommand{\R}{\mathbb{R}}
\renewcommand{\P}{\mathcal{P}}
\renewcommand{\H}{\mathcal{H}}
\newcommand{\1}{\mathbf{1}}
\newcommand{\D}{\mathcal{D}}
\renewcommand{\v}{\mathbf{v}}
\renewcommand{\u}{\mathbf{u}^1}
\newcommand{\m}{\mathbf{u}^2}
\newcommand{\e}{\mathbf{u}}
\newcommand{\bt}{\boldsymbol{\theta}}
\newcommand{\be}{\boldsymbol{\eta}}

\newtheorem{theorem}{Theorem}
\newtheorem{assumption}{Assumption}
\newtheorem{definition}{Definition}
\newtheorem{lemma}{Lemma}

%% file: intro.tex
\section{Introduction}
\label{section:intro}

Recommender systems are AI-driven systems that have come to influence nearly every aspect of human activity on the internet and, most importantly, shape the information and opportunities that are available to us. This includes, for instance, the news we read, the job listings we are matched to, the entertainment we consume, and the products we purchase. Due to this influence on modern life, it has become crucial to ensure that recommender systems are \emph{aligned} with the goals and values of their users and society at large. 
However, it is well documented that current recommender systems do not always succeed at alignment \citep{Stray+21, stray2022building}.
For example, there is evidence that a 
fraction of the time spent by users on online platforms can be attributed to impulsive usage  \citep{gruning2023directing, allcott2022digital, cho2021reflect}. 
It has been also been observed that 
recommendation algorithms lead users into narrower selection of content over time which lacks diversity and results in echo-chambers \citep{CarrollD0H22, JiangCLGK19, kalimeris2021preference}. 
Moreover, several works have studied the prevalence of  problematic content such as conspiracy theories, hate speech and other radical/violent content on  recommendation platforms \citep{faddoul2020longitudinal, ribeiro2020auditing, ledwich2019algorithmic, stray2023algorithmic}.

However, aligning recommenders with the goals of users is challenging because the utility/preferences that operationalize these goals are not known explicitly.
Most recommendation platforms use 
engagement signals such as likes, shares, watch time etc.,
as a proxy for utility and optimize content 
for users based on these signals. Such signals are abundantly available and one can
train machine learning models to predict these signals with high accuracy.
However, there is a wealth of literature 
grounded in established human psychology 
which suggests that engagement signals of users are not always aligned with their utility \citep{LyngsBKS18, LyngsLSBSIKS19, MilliBH21, KleinbergMR22}. 
\cite{KleinbergMR22} explain this misalignment by considering a ``dual system'' model for human decision-making:
a swift and impulsive \emph{System-1} whose decisions are driven by short-term satisfaction, and a logical and forward-looking \emph{System-2} which 
makes decisions according to the utilities and long-term goals of the user\footnote{Note that this ``dual system'' model is a simplification or abstraction of specific psychological mechanisms and our usage of the terms System-1/2 might deviate slightly from their usage in the psychology literature \citep{thaler1981economic, akerlof1991procrastination, laibson1997golden}.  }.
Due to its impulsiveness and short-term orientation, System-1 behavior is susceptible to engagement that may not be aligned with a user's utility, e.g., content such as click-bait, toxic, or low-information content.
\cite{KleinbergMR22} consider a specific 
model for user interaction during a single session where System-2 decisions are confounded by System-1 decisions and the platform only observes the engagement signals for each session which are a combination of the two decision processes.
Using this model, they show that it is difficult to disentangle between System-1 and System-2 behavior using this session-level engagement signal. 
For example, users might continue scrolling their feed impulsively beyond the limits dictated by their utility but it is difficult for the platform to tell whether some of the usage was driven by System-1.
Moreover, \cite{KleinbergMR22} show that optimizing recommendations based on such engagement signals can cause users to quit the platform entirely. This is because it can 
lead the user towards more and more System-1 usage which can lead to longer sessions with lower overall utility than other options outside the platform. 
In this work we ask the following questions:
\begin{itemize}
\item \emph{Are there signals that are better aligned with user utility than engagement signals such as likes, shares, watch time?}

\item \emph{Can we use this signal to estimate user utility and ultimately recommend content based on it?}
\end{itemize}

In this paper we explore a new approach to recommender systems where we infer user’s utility based on their return probability to the platform rather than engagement signals. 
Our approach is based on 
the intuition that users tend to return to a platform in the long run if it creates utility for them, while pure engagement-driven interactions, i.e., interactions that do not add meaningful utility, may affect user return in the short term but will not have a lasting effect. 
Hence, instead of trying to estimate utility from engagement signals (which are susceptible to be driven by System-1 behavior), we focus on modeling the probability that a user will keep returning to the platform in the long-run (which is more likely to be a conscious System-2 decision).

To this end, we propose a generative model of user arrival rates based on a self-exciting Hawkes process where the probability to return to the platform depends on their experiences during past sessions. 
There is substantial empirical evidence which suggests that a user's return to the platform can also be driven by impulsive behavior in addition to utility-driven behavior 
 \citep{cho2021reflect, moser2020impulse, lyngs2019self}. 
Hence, inspired by the ``dual system'' model of \cite{KleinbergMR22}, we consider the influence of both System-1 and System-2 decision-processes in modeling user return probability.   
In particular, the triggering kernel of the Hawkes process has two components: \textbf{1)} A trigger intensity based on a \emph{System-1} process driven by the instantaneous gratification or \emph{moreishness} from past interactions with a rapidly vanishing effect of the user's return probability; \textbf{2)} A trigger intensity based on \emph{System-2} process driven by the utility of past interactions and with a more steady longer-term impact on the user's return probability. Our model allows for the possibility that a user keeps returning to the platform even when the 
platform always recommends moreish content 
because maybe the System-1 trigger is high enough for the user to keep returning (even though it lasts for a short while).  
However, we do not model long-term addictive behavior where the presence/absence of the user might not correlated with their experience on the platform.  

Given past user sessions, our goal is then 
 to learn \emph{disentangled} representations of user's impulsive and utility preferences that parametrize this model of return probabilities. 
By disentangled, we mean that we learn two representation for each user that capture their utility and moreishness. 
Given properly disentangled representations, it is then possible to shift from engagement-based recommendations to strategies that are better aligned with user's utility.

For this purpose, we make the following 
contributions
\begin{itemize}
\item Our main theoretical result is to show that under mild
identifiability conditions, the two different components 
of the trigger intensities are \emph{uniquely identifiable using maximum likelihood estimation}.
This allows us to identify System-1 and System-2 behavior from the observed user interactions. 
See Section~\ref{sec:identifiability} for more details.
\item Experimentally, we show on synthetic data that (a) we are able to infer disentangled representations from purely observational data and that (b) that optimizing recommendations based on the estimated user's utility, largely increases their utility compared to engagement-based systems.
See Section~\ref{sec:expts} for more details.
\end{itemize}

%% file: problem.tex
\section{Preliminaries and Problem Setup}
\label{sec:problem}

 \subsection{Dual System and Inconsistent Preferences}

We first ground our discussion in the dual systems
theory that is based on established psychology mechanisms \citep{kahneman2011thinking}.
The dual systems theory proposes two different 
processes for human decision-making: (1) a swift, parallel and intuitive \emph{System-1};
(2) a slow, logical, and long-term oriented 
\emph{System-2}.
The System-1 responses are driven by short-term satisfaction and it allows tasks to be performed instinctively without conscious awareness.
For example, responses like quickly picking up one's phone 
to check for notifications constitute System-1 responses.
The System-2 responses are driven by long-term 
goals and require logical decision-making and planning. 
For example, responses like performing a statistical analysis or 
watching videos to improve your skiing technique constitute System-2 responses.

The dual systems theory gives interesting insights 
in the context of usage of online platforms. 
Based on this theory, several papers~\citep{LyngsLSBSIKS19, LyngsBKS18, MilliBH21} have argued that engagement signals may be more correlated with System-1 responses than System-2's, with the risk of leading to recommendation strategies optimized for impulsive behavior rather than user's utility.
\cite{KleinbergMR22} formalized this scenario by 
considering a simple model for user interaction within a single session.
Under this model, the net \emph{value} of the user is the \emph{utility} derived from System-2-based use 
of the platform (e.g., reading useful news articles) minus the \emph{utility from an outside option} lost 
due to System-1 (impulsive) usage of the platform (e.g., spending time on click-bait news).
As the engagement signal (e.g., the total reading time) is possibly the result of both System-1 and System-2 responses, it is not possible to estimate the actual utility obtained by the user using this signal.

If the platform optimizes for engagement as a proxy 
for value, then this might lead to recommending moreish content.
This will hurt the overall long-term utility of users since it will increase System-1 usage of the platform (which is further increased by the feedback loop of engagement and recommendation).
\cite{KleinbergMR22} considered a simple model for 
user arrival-- the user will arrive to the platform as
long as it results in positive net value.
As soon as the System-1 behavior exceeds a certain threshold, the user will quit the platform due to insufficient utility.
In this case the platform might only realize about this excess System-1 usage when the user has already quit.
Hence, it can be difficult to isolate System-1 
behavior from System-2 behavior while the user is 
still present on the platform.

However, this model for user arrival is unsatisfactory as 
it suggests a binary choice in terms of user arrival rates-- either the user arrives because of positive value or does not arrive because of negative value.
In this work we consider a model where the user arrival rates 
depend on the utility and can increase/decrease based on user satisfaction.
In the next section we define temporal point processes which will be used to model arrival rates of individual users.

\subsection{Temporal Point Processes and The Hawkes Process}
\label{sec:tpp}
A temporal point process (TPP) is a stochastic process whose realization
is a sequence of events $\H = \{t_i\}_{i=1}^k$ 
where $t_i$ is the arrival time of the $i$-th event.
We will denote by $\H_{t^-} = \{t_i \in \H : t_i < t\}$ the set of 
event arrival times up to but not including $t$.
There are two major approaches for describing a TPP.
The first approach is to model the distribution of interevent times, i.e.,
the time lengths between subsequent events. 
Given history $\H_{t^-}$, we denote by $f(t \vert \H_{t^-})$
the conditional density function of the time of the 
next event. 
The joint density of the distribution 
of all events is given by 
\[
f(t_1, \cdots, t_n) = \prod_{i \in [n]} f(t_i \vert \H_{t_{i}^-})
\,.
\]
A popular approach for describing a TPP is through 
the conditional intensity function (or hazard function)
$\lambda(t)$:  
\[
\lambda(t) = \frac{f(t\vert \H_{t^-})}{1-F(t \vert \H_{t^-} )}
    \,,
\]
where $F(t \vert \H_{t^-})$ is the cumulative conditional
density function.
It can be shown that 
\[
\lambda(t) \mathrm{d} t = \Ex[N (t, t+ \mathrm{d}t) \vert \H_{t^-}]
\,.
\]
where $N(t_1, t_2) = \sum_{t \in \H} \1[t \in [t_1, t_2)]$ is the counting process.
Hence, the expected number of arrivals in a time-interval $[t_1, t_2]$ is given by $\int_{t_1}^{t_2} \lambda(t) \mathrm{d}t $. 
If we model  user arrivals as TPPs, we can  
estimate quantities such as expected number of sessions per day per user and expected number of active users per day, by estimating the underlying intensity functions. 

Given a sample $\H$ from the point process over a time-horizon $T$, the likelihood function is defined as
\[
L(\H) = \left(\prod_{i=1}^k  \lambda(t_i)\right) \exp\left(- \int_{0}^T \lambda(t) \mathrm{d} t \right)
    \,.
\]
One can maximize this function  to estimate the intensity function or parameters that govern the intensity function.
In some cases, the point process is such that each arrival is associated with a special mark/feature, for example, each earthquake is associated with a magnitude.
We refer to such processes as marked point process.
The conditional intensity function for the marked 
case is then given by $\lambda(t, \kappa) = \lambda(t) g(\kappa | t ) $
where $g(\kappa | t)$ is the conditional density of the mark distribution.
If the goal is also to jointly learn the parameters of the mark distribution along with the parameters of the TPP, then we include the density of the mark distribution in the likelihood computation.

Hawkes processes are a special class of temporal point 
processes where the intensity at  any given time 
is influenced by past arrivals.
Hawkes processes are also referred to as self-exciting point process.
Specifically, a Hawkes process with exponential decay is  
defined according to a conditional intensity 
function
\begin{equation*}
\lambda(t) = \mu + {\sum_{t' \in \H_{t-}}} \alpha \beta e^{-\beta (t-t')} \,,
\end{equation*}
where $\mu > 0$ is the base intensity, 
$\alpha > 0$ is the infectivity rate, i.e., the expected
number of events triggered by any given event, 
and $\beta > 0$ is the decay rate. 
If the infectivity rate $\alpha$ is $0$ then we recover the Poisson process.
For the Hawkes process with exponential decay,
one can calculate the likelihood function efficiently 
without the need to perform Monte Carlo estimation 
to evaluate the integral.
In the case of marked Hawkes process, we have
\begin{equation*}
\lambda(t) = \mu + {\sum_{t' \in \H_{t-}}} \alpha_{t'} \beta e^{-\beta (t-t')} \,,
\end{equation*}
where the infectivity rate $\alpha_t$ has a dependence on the arrival time $t'$ but not on the history.

\subsection{Our Recommender Model}

We consider the interaction of a recommendation
platform with a population of $[m]$ users and $[n]$ items. Each item $j \in [n]$ is represented by a single embedding $\v_j \in \R^d$ which represents the items latent features. Furthermore, each user $i \in [m]$ is represented by \emph{two} embeddings $\u_i \in \R^d$ and $\m_i \in \R^d$ which represent the user's System-1 and System-2 characteristics  corresponding to 
moreishness and utility, respectively. 
We will further assume that the embeddings are normalized such that $\| \u_i\|_2 \leq 1$ and $\|\m_i\|_2 \leq 1$.
Whether an item is then aligned with a user's preferences with regard to moreishness (impulsiveness)
or utility (long-term goals), is then modeled via the inner products
\begin{align*}
    \text{Moreishness: }  \v_j^\top \u_i \quad\quad
    \text{Utility: }  \v_j^\top \m_j
\end{align*}

In the following, we assume item embeddings $\v_j$ are known since there is abundant 
data available that describes each item, for example, item attributes,  audio-visual features and engagement signals.\footnote{In our setup, we only use engagement signals for learning item embeddings and rely solely on arrival rates for learning user embeddings.}
However, we assume user embeddings are unknown and our goal is to infer them from past content interactions such that System-1 and System-2 characteristics are disentangled.
For this purpose, our model incorporates 
the dual system theory into the 
\emph{arrival process} of users to the platform such that the probability to arrive at the platform is governed by both System-1 and System-2 decision-processes. This allows us to connect past content interaction with the long-term behavior of users, i.e., their return to the platform, and as such obtain the necessary signal to disentangle System-1 and System-2 charactersitics.

Formally, the arrival of user $i \in [m]$ to the 
platform is governed by a Hawkes process with 
the conditional intensity function $\lambda_i(t)$, defined as
\begin{align}
\label{eq:intensity}
\lambda_i(t) = \mu_i + {\sum_{t' \in \H_{it^-}}} \alpha^1_{it'} \beta^1_i \cdot e^{-\beta^1_i( t-t')} + \alpha^2_{it'} \beta^2_i \cdot e^{-\beta^2_i( t-t')}
    \,,
\end{align}
where $\mu_i$ is the base intensity, $\H_{it^-} $ is the history of 
past arrival times of user $i$ up to but not including time $t$,
$\alpha^1_{i,t'}$  and $\alpha^2_{i,t'}$ are System-1 and 2 infectivity rate, respectively, 
and, $\beta^1_i$ and $\beta^2_i$ are System-1 and 2
decay rates\footnote{In this work we do not consider a dependence between the arrival rates of different users, instead we focus on isolating System-1 and System-2 effects.}\footnote{It is possible to further assume that the process has finite memory and it only depends on the recent history. }.
We also assume that $0\leq \beta^2_i < \beta^1_i$, $\forall i$. 
This assumption implies that System-1 intensity decays faster than System-2's.
The justification for this assumption is that \emph{utility driven
sessions drive sessions in the long-term}. In contrast, moreishness 
driven sessions influence sessions only in the short-term as users might get bored if the usage is being driven purely by interaction/engagement and not utility.
Hence, System-1 interactions contribute only a short spike in arrival intensity, while System-2 interactions contribute longer-lasting effects.

Next, when user $i$ arrives at time $t$, the platform
recommends a set $S_{i,t}$ of items. 
We will denote by $S_{i,t} = \{s_{i,j,t}\}_{j =1}^{ l_{i,t}}$ the set of items that user $i$ interact within the session corresponding to time $t$.
We will denote by $\v_S$ the vector summarizing a session $S$. In particular, we let $\v_S :=1 / |S| \sum_{j\in S} \v_i$.
Note that, unlike \cite{KleinbergMR22}, we do not assume that 
sessions are generated according to a particular stochastic 
model and our focus is on modeling the arrival rates instead. 
Hence, we do not assume a model for how the engagement signal is generated, but our understanding is that both $\u$ and $\m$ combine together in some way to generate the engagement signal.

Given a user session, we can then model its contribution to the arrival intensity via its infectivity rate. In particular, the System-1 and System-2 infectivity rates $\alpha^1_{i,t}$ and $\alpha^2_{i,t}$ 
are defined as
\begin{equation}
\label{eq:alpha}
\alpha^1_{it} = \phi(\v_{S_{it}}^\top \e_i^1), \qquad
\alpha^2_{it} = \phi(\v_{S_{it}}^\top \e_i^2)
    \,,
\end{equation}
where  $\phi: \R \rightarrow [0,0.5]$ is a link function.\footnote{
We remind the reader that the goal of our modeling assumptions is to abstract away the 
details of user interaction 
that are not important in to model long-term value.
For example, the user session might also involve activities 
other than scrolling through the recommendations. 
The users might also be inclined to spend more time in the session
if the realized recommendations happen to be more aligned 
with their interests.}
We let the range of  $\phi$ to
be $[0,0.5]$ because we need to ensure that $0 \leq \alpha_{it}^1 + \alpha_{it}^2 \leq 1$ so that the underlying Hawkes process is \emph{stationary} and \emph{ergodic} \citep{Guo+18}.

\subsection{Goal}
Equations (\ref{eq:intensity}) and (\ref{eq:alpha}) connect the return probabilities and content interactions of users with their System-1 and System-2 characteristics. 
Given past interactions $\D_i = \{(t, S_{i,t})\}$ for a user $i \in [m]$, our goal is then to learn user representations $\u_i, \m_i$ from $\D_i$. 
As we will show in \cref{sec:identifiability}, incorporating the temporal signal of return probabililties into the inference process allows us then to disentangle the effect of System-1 and System-2 decision-processes.
In addition to user embeddings, we also need to estimate the (nuisance) parameters $\mu_i, \beta^1_i, \beta^2_i$ for each user $i \in [m]$ using the observed interactions since they are central for an accurate disentanglement of System-1 and System-2 behavior. 
Note that the problem of learning for each user can be solved independently because of the assumption that 
the item embeddings are known and the point process for different 
users don't influence each other.
We contrast our approach with matrix factorization where the item and user embeddings are jointly learnt and data from multiple users is pooled together.\footnote{In our setting, if the platform suspects that behavior of multiple 
users is very similar and it would be beneficial to pool the data 
together from these users, then one can create a super user 
with all the data pooled together and learn a joint embedding 
for this super user which can be refined subsequently.}

Once we estimate $\e^2_i$ we can rank items according to their utility 
by taking the dot-product of the corresponding item-embedding with $\e^2_i$.
Hence, given this estimate of $\e^2_i$
one can \emph{maximize per-session utility}
by recommending items that maximize the 
dot product to $\e^2_i$. In particular, for deterministic ranking, an item is ranked at location $R_i$ via
\begin{equation}
\text{Deterministic ranking:}\quad R_i = \operatorname{arg\,sort}_{\v_j} \langle \m_i, \v_j\rangle
\label{eq:u2ranking}
\end{equation}
i.e., \emph{only} via its System-2 representation. For stochastic rankings, a similar approach can be used using a temperature controlled softmax function. 
We defer the discussion on other platform objectives to Section~\ref{sec:discuss}.
The next section delves into the identifiability of these parameters
and shows that \emph{utility can be disentangled from moreishness} under our model.

%% file: identifiability.tex
\section{Identifiability and Consistency}
\label{sec:identifiability}

In order to optimize content with respect to utility via recommendations such as in \cref{eq:u2ranking}, one needs to \emph{reliably} disentangle utility from moreishness. 
However, this is a non-trivial task as it is not immediately clear if samples from the underlying Hawkes 
process are enough to identify the two different components of the trigger intensity. 
The core challenge here is the model does not only need to correctly infer \(\u_i, \m_i\), but also the parameters \(\beta_i^1, \beta_i^2, \mu_i\) which all influence the intensity function. 
Moreover, identifiability results for Hawkes processes 
are mainly known for settings where infectivity rates $\alpha$'s are stationary and do not vary with time \citep{Guo+18}.  
Under what conditions can we assume that we can infer these parameters accurately from past interactions $\D_i$?
In the following, we show that it indeed possible to disentangle System-1 and System-2 behavior in our model and 
thereby enabling content optimization with respect to utility. 
For this purpose, we establish the identifiability of model parameters 
and show that maximum likelihood estimation (MLE)
leads to a consistent estimator.
We consider a single user in this discussion.
We start with a definition of identifiability for statistical models.
\begin{definition}[Identifiability]
A class of statistical models $\P = \{P_{\bt}: \bt \in \Theta \}$ is said to be \emph{identifiable} if $P_{\bt_1}$ = $P_{\bt_2}$  implies that $\bt_1 = \bt_2$ for all $\bt_1, \bt_2 \in \Theta$.
\end{definition}

We will now use the following sufficient condition for the 
identifiability  of Hawkes processes \citep{Guo+18}.
Let us denote by $\kappa(t)$ the trigger intensity of the Hawkes process,
i.e., $\kappa(t)$ is such that $\lambda(t) = \mu + \sum_{t' \in \H_{t^-}} \kappa(t - t')$. Also, let $\be$ be the set of parameters that govern the trigger intensity $\kappa$.
We will use $\kappa(t; \be)$ to make the dependence on parameters $\be$ explicit.
Let $\bt = (\mu, \be)$ be the set of all parameters that govern the intensity $\lambda$. 
\begin{lemma}[\cite{Guo+18}]
A class of Hawkes processes $\{\lambda(t; \bt): \bt \in \Theta\}$ is identifiable if the corresponding trigger intensity $\kappa$ is identifiable, i.e.,
if $\kappa(t;\be_1) = \kappa(t; \be_2)$ $\forall t$, then  $\be_1 = \be_2$.
\end{lemma}
The above lemma allows us to establish the identifiability 
of the Hawkes process by proving the identifiability of 
the corresponding trigger intensity.
We will now focus on the identifiability of the trigger intensity $\kappa$.
We now mention the technical 
assumptions that are required to prove our result.

\begin{assumption}
\label{assum:1}
We assume that 
$\alpha^1_t =  ( \u)^\top f\left( \v_{S_t} \right)  + c^1$
and
$\alpha^2_t =  (\m)^\top f\left( \v_{S_t} \right)  + c^2$
where $\m \in \R^d$ is the (unknown) user embedding for utility, $\u \in \R^d$ is the 
(unknown) user embedding for moreishness, $S_t$ denotes the session at time $t$ 
and $\v_{S_t} \in \R^d$ is the (known) session vector, and the (known) normalizing function $f: \R^d \rightarrow \R^d $ and constants $c^1, c^2$ are such that they ensure $0 \leq \alpha_t^1, \alpha_t^1 \leq 0.5$. 
\end{assumption}
This assumption is satisfied when the link 
function $\phi(x) := (x+1)/4$  
in \cref{eq:alpha}, and the function $f$ performs $\ell_2$ normalization because $\phi$ has range $[0,0.5]$ due to $\|\u\|_2, \|\m\|_2 \leq 1$.
Note that $f(\v_{S_t})$ is known 
because the session $S_t$ and the corresponding item embeddings are known.
This assumption is  for simplicity of analysis and we believe that our identifiability results will hold as long as $\phi$ is a one-to-one mapping.
\begin{assumption}
\label{assum:2}
Session $S_t$ is \emph{deterministic} given time $t$ and does not depend on the realization of arrival times $\H$. 
\end{assumption}
This assumption is crucial for identifiability and consistency because if the session can depend on previous realizations of arrival times, then $\v_{S_t}$ becomes a random variable that can influence future arrivals.
Note that
the session can still depend on the parameters $\e^1$ and $\e^2$, but it cannot be dependent on the arrival times. 
We only make the assumption of \emph{determinism} to simplify the 
presentation and one can allow for randomness in session 
generation that is independent of the previous arrival times.
Under the assumptions above, the trigger function can be written as 
\begin{align}
\label{eq:trigger}
\kappa(t) 
= \Big(\e^1 \beta^1\exp(-\beta^1t) + \e^2 \beta^2 \exp(-\beta^2 t) \Big)^\top f(\v_{S_t})  \\
\qquad + \Big(c^1 \beta^1 \exp(-\beta^1t) + c^2 \beta^2 \exp(-\beta^2 t) \Big)
    \,.
\end{align}

\begin{assumption}
\label{assum:3}
For each vector $\e \in \R^d$, there exists some time $t > 0$ such that $\e^\top f(\v_{S_t}) \neq  0$. In other words, the set of vectors $\{f(\v_{S_t})\}_{t = 0}^\infty$ span the entire 
$d$-dimensional Euclidean space.
Moreover, there exists an interval $[a_1,a_2] \subseteq [0,T]$ where $S_t$ remains fixed over $t \in [a_1,a_2]$. 
\end{assumption}
The above assumption is not much stronger than the 
assumption required for complete recoverability in linear regression.
In other words, we need to span the entire $d$-dimensional Euclidean space using vectors $\{f(\v_{S_t})\}_{t = 0}^\infty$ because we need to recover $\u,\m$ by measuring their dot-products with each $f(\v_{S_t})$.
The additional assumption about $S_t$ remaining fixed during a small interval requires that the user 
will see the same set of items regardless of when the user arrives at the platform within this interval. This can happen in practical settings, for instance, 
when the two log-in events are sufficiently close that the 
feed does not refresh. This assumption is used for the 
identifiability of $\beta$'s.
Note that we only require one such interval to exist.

Also, note that our results apply for a non-stationary set of items. We only make the assumption of fixed set of items 
for the sake of convenience. 
We are now ready to show that the above trigger function is identifiable. 
\begin{theorem}
\label{thm:identifiability}
Under Assumptions~\ref{assum:1},~\ref{assum:2} and~\ref{assum:3},
the trigger function in Equation~\ref{eq:trigger} defined over the domain $\R_+$ is identifiable if $\beta^1\neq \beta^2$, $\beta^1, \beta^2 > 0$, $\|\e^1\|_2 $, $\|\e^2\|_2 <1$ and $f: \R^d \rightarrow \R^d$ is a known function. 
\end{theorem}

\begin{proof}
We want to show that given a fixed (bounded) function $g(t) = f(\v_{S_t})$, 
there is a unique set of values $\e^1, \e^2, \beta^1, \beta^2$ 
that generate a given trigger function $\kappa(t)$.
Suppose for the sake of contradiction that there are two different set of values $\be_1 = (\e_1^1, \e^2_1, \beta^1_1, \beta^2_1)$ 
and $\be_2 = (\e^1_2, \e^2_2, \beta^1_2, \beta^2_2)$ that generate the 
same trigger function $\kappa(t)$, i.e., 
$\kappa(t; \be_1) = \kappa(t; \be_2)$ for all 
measurable sets in the domain.

We first show that $\beta_1^1 = \beta_2^1$ and $\beta^2_1 = \beta^2_2$.  
According to Assumption~\ref{assum:3}, there exists an interval $[a_1, a_2]$ such that $S_t$ is fixed over this interval.
Then we have that $\forall t \in [a_1, a_2]$, 
with $S_t = S$, the trigger intensity can be written as
\begin{align*}
\kappa(t; \be_1) 
= \alpha^1 \beta^1_1 \exp(-\beta^1_1 t) + \alpha^2 \beta^2_1 \exp(-\beta^2_1 t) \\
\kappa(t; \be_2) 
= \alpha^1 \beta^1_2 \exp(-\beta^1_2 t) + \alpha^2 \beta^2_2 \exp(-\beta^2_2 t)
    \,,
\end{align*}
where $\alpha^1 = (\e^1)^\top f(\v_{S}) + c^1$
and $\alpha^2 = (\e^2)^\top f(\v_{S}) + c^2$.
Hence, the $\alpha$'s remain fixed over the time interval $[a_1, a_2]$. Since, the above trigger intensity takes 
the form of sum of exponential functions, it easy to establish using Lemma~\ref{lem:trigger_1} that  
$\kappa(t; \be_1) = \kappa(t; \be_2)$ for all $t \in [a_1, a_2]$ implies that $\beta_1^1 = \beta_2^1$ and $\beta^2_1 = \beta^2_2$.

Now, given $\beta^1 = \beta_1^1 = \beta_2^1$ and $\beta^2 = \beta^2_1 = \beta^2_2$, we have, $\forall t>0$,
\begin{align*}
\kappa(t; \be_1) - \kappa(t; \be_2) 
& = ((\e^1_1 - \e^1_2) \beta^1\exp(-\beta^1 t) \\
 &\qquad+ (\e^2_1 -\e^2_2)\beta^2 \exp(-\beta^2 t) )^\top f(\v_{S_t})  \\
 	&= 0
    \,.
\end{align*}
Note that the vector $(\e^1_1 - \e^1_2) \beta^1\exp(-\beta^1 t) + (\e^2_1 -\e^2_2)\beta^2 \exp(-\beta^2 t)$  is a linear combination of two vectors $(\e^1_1 - \e^1_2)$ and $(\e^2_1 -\e^2_2)$.
If the dot product with $f(\v_{S_t})$ is $0$ for all $t$, then it could only mean that 
$(\e^1_1 - \e^1_2) \beta^1\exp(-\beta^1t) + (\e^2_1 -\e^2_2)\beta^2 \exp(-\beta^2 t) = 0$.
Hence, we have that $(u^1_{1,i} - u^1_{2,i}) \beta^1\exp(-\beta^1t) + (u^2_{1,i} -u^2_{2,i})\beta^2 \exp(-\beta^2 t) = 0$.

Now,  consider the following function 
$\kappa'(t) =  u^1 \beta^1\exp(-\beta^1t) + u^2 \beta^2  \exp(-\beta^2 t)$.
Since
$\kappa(t; \eta_1) = \kappa(t; \eta_2)$ and 
$g(t) > 0$ is a deterministic function,
we have that $\kappa'(t; \eta_1) = \kappa'(t; \eta_2)$.
This implies that $\kappa'$ is not identifiable,
which is a contradiction according to Lemma~\ref{lem:trigger_1} in the Appendix.
\end{proof}

We now shift our focus to consistency. 
\begin{definition}[Consistency]
We say that a parameter estimation procedure for a 
class of statistical models $\P = \{P_{\bt}: \bt \in \Theta \}$ is consistent if the estimate $\hat{\bt}_k$ given $k$ samples from $P_{\bt}$ satisfies $\hat{\bt}_k \rightarrow \bt$
as $k \rightarrow \infty$. 
\end{definition}
We utilize the results of \cite{Guo+18} to establish the consistency of maximum likelihood estimation (MLE) under our model. \cite{Guo+18} identify a set of technical conditions on the underlying Hawkes process that
ensure consistency of MLE.
The main condition amongst these is identifiability of the model which is satisfied because of Theorem~\ref{thm:identifiability}. 
The second condition of stationarity is ensured by our 
model due to the fact that $\alpha_t^1+\alpha_t^2 < 1$, $\forall t$.
The final condition is the compactness of $\Theta$ which is easily satisfied by our condition on the parameter space.
The following theorem gives our result.
\begin{theorem}\label{thm:consistency}
Under Assumptions~\ref{assum:1},~\ref{assum:2} and~\ref{assum:3},
the MLE of our Hawkes process 
recommender model is consistent. 
\end{theorem}

Theorems \ref{thm:identifiability} and \ref{thm:consistency} together state that we can identify all the parameters defining the overall recommendation process by just observing samples generated by its associated Hawkes process. Crucially, while observations of engagement confound moreishness and utility, the return process allows us to discriminate between these two components.

%% file: expts.tex
\section{Experiments}
\label{sec:expts}
We perform experiments on synthetically generated data
to demonstrate the usefulness of our approach for maximizing user utilities. 
We generate user interactions according to our model
with known parameters. 
We evaluate our approach under two broad metrics: (1) how well it can recover the 
underlying model parameters, (2) how much better can we do in terms of utility maximization 
using our approach as compared to engagement optimization.

Since our problem can be solved independently for each user, we run experiments from the perspective of a single user.
We set the number of items $m = 1000$, and 
the embedding dimension $d = 10$.
We use the link function $\phi: [-1,1] \rightarrow [0,0.5]$ defined as $\phi(x) := (x+1)/4$ in order to define the infectivity rates $\alpha^1$ and $\alpha^2$ in Equation~\ref{eq:intensity}.
We assume that the embedding vectors $\e^1,$ $\e^2$
as well as the item vectors $\v$'s have a unit $\ell_2$ norm,
which ensures that their dot products are in the interval $[-1,1]$. 
We set Hawkes process 
parameter values as follows: $\mu = 0.3$, 
$\beta^1 = 4$ and $\beta^2 = 1$ (if not stated otherwise), so that the utility-driven System-2 process has a long-lasting effect on the return probability compared to moreishness-driven System-1 behavior.
We generate user session arrival times according to the underlying Hawkes process
using the well-known thinning algorithm \citep{ogata1981lewis}.
We first generate the number of items in a user session according 
to a geometric random variable  that is clipped to be in the range $[1,6]$ with probability of heads $p = 0.8$. Once we fix the number of items, each item is selected randomly from the set of available items, thus ensuring that the set of items is well covered, coherently with Asm.~\ref{assum:3}.
We divide the entire sequence of sessions into several epochs, where each epoch contains $1000$ sessions.
This is useful in order to reduce the computation complexity of 
log-likelihood computation which is quadratic in the number of sessions.
We treat each of these epochs as a \emph{separate sample} from the underlying Hawkes process.
The log-likelihood are then maximized using stochastic gradients  
with mini-batching.
We do not need to use the log-likelihood 
for the marked case as we are not interested in learning the parameters of the mark distribution (see ~\cref{sec:tpp}). 
We use Adam with a uniform learning rate of
$0.002$ and a batch size of $16$.

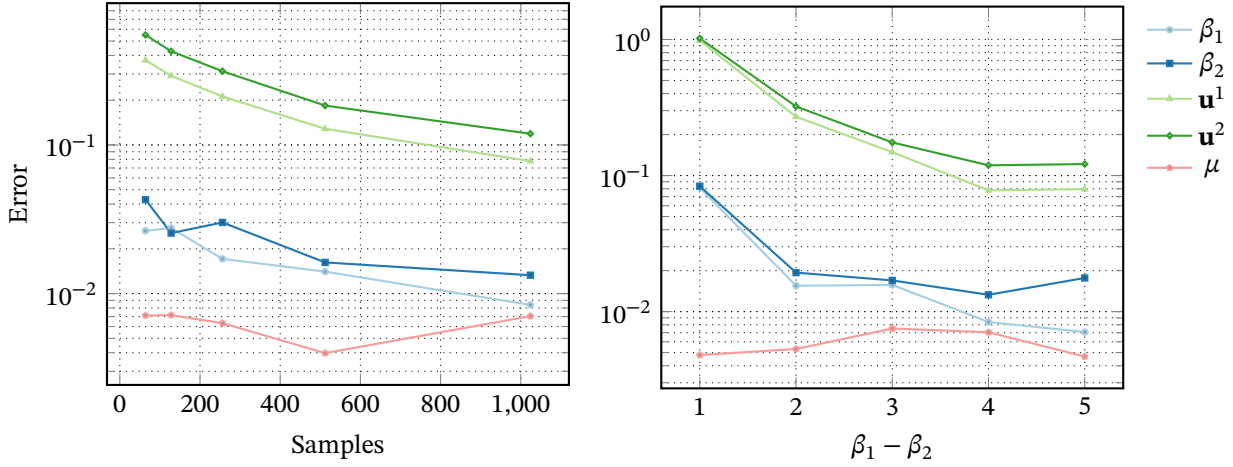
\begin{figure}[t]
    \centering
    \hspace{-3.5em}%
    \begin{subfigure}{0.49\linewidth}
        \centering
        \begin{tikzpicture}
            \begin{axis}[
                ymode=log,
                grid=both,
                thick,
                mark size=1pt,
                width=.95\textwidth,
                xlabel=Samples,
                ylabel=Error,
                every axis plot post/.append style={%
                    error bars/.cd, 
                    y dir=both, 
                    y explicit, 
                    error bar style={line width=1pt},
                },
            ]
                \addplot table [col sep=comma, x={samples}, y=beta_1_err] {./dat/error_vs_samples.csv};
                \addplot table [col sep=comma, x={samples}, y=beta_2_err] {./dat/error_vs_samples.csv};
                \addplot table [col sep=comma, x={samples}, y=u_1_err] {./dat/error_vs_samples.csv};
                \addplot table [col sep=comma, x={samples}, y=u_2_err] {./dat/error_vs_samples.csv};
                \addplot table [col sep=comma, x={samples}, y=mu_err] {./dat/error_vs_samples.csv};
            \end{axis}
        \end{tikzpicture}
    \end{subfigure}%
    \begin{subfigure}{0.49\linewidth}
        \centering
        \begin{tikzpicture}
            \begin{axis}[
                ymode=log,
                grid=both,
                thick,
                mark size=1pt,
                width=.95\textwidth,
                xlabel=\(\beta_1 - \beta_2\),
                legend pos=outer north east,
                legend style={draw=none},
                every axis plot post/.append style={error bars/.cd, y dir=both, y explicit, error bar style={line width=1pt}},
            ]
                \addplot table [col sep=comma, x={beta_1 - beta_2}, y=beta_1_err] {./dat/error_vs_beta_gap.csv};
                \addplot table [col sep=comma, x={beta_1 - beta_2}, y=beta_2_err] {./dat/error_vs_beta_gap.csv};
                \addplot table [col sep=comma, x={beta_1 - beta_2}, y=u_1_err] {./dat/error_vs_beta_gap.csv};
                \addplot table [col sep=comma, x={beta_1 - beta_2}, y=u_2_err] {./dat/error_vs_beta_gap.csv};
                \addplot table [col sep=comma, x={beta_1 - beta_2}, y=mu_err] {./dat/error_vs_beta_gap.csv};
                \legend{$\beta_1$, $\beta_2$, $\u$, $\m$, $\mu$};
            \end{axis}
        \end{tikzpicture}
    \end{subfigure}
     \caption{The error in parameter estimation as a function of number of samples on the left and gap in decay rates on the right. }
    \label{fig:error}
\end{figure}

\subsection{Effect of sample size on the estimation error}
\label{sec:error}
In this experiment, our goal is to demonstrate that our algorithm is 
able to learn the parameters of the Hawkes process as well as user embeddings given sufficient number of samples.
We generate item embeddings as follows:
(1) generate a random matrix $A$ of size $d \times d$, (2) compute a QR factorization of $A$, i.e. $A = Q \times R$, (3) generate $1000$  
item embeddings with dimension $d$ by taking a random row of $Q$ and adding $\mathcal{N}(0,1/10 d)$ noise to each dimension, (4) each vector is normalized to have norm $1$. 
We let $\e^1 = Q_1$ and $\e^2 = Q_2$ where $Q_1$  and $Q_2$ are the first and second row of $Q$, respectively.
This ensures that the two embeddings are orthonormal.
We report the $\ell_2$ norm
of the distance between the estimates and the true values divided by the $\ell_2$ norm of the true values.
For scalars, this is just the percentage error in terms of absolute values.
Figure~\ref{fig:error} reports the error in estimation as
a function of the number of samples from the Hawkes process (each sample has $1000$ sessions) that were used for learning.
One can observe that the error reduces as a function of the number of samples.
Also, the model clearly identifies the 
two different components of the Hawkes process
and is able to disentangle System-1 behavior and  
System-2 behavior correctly.

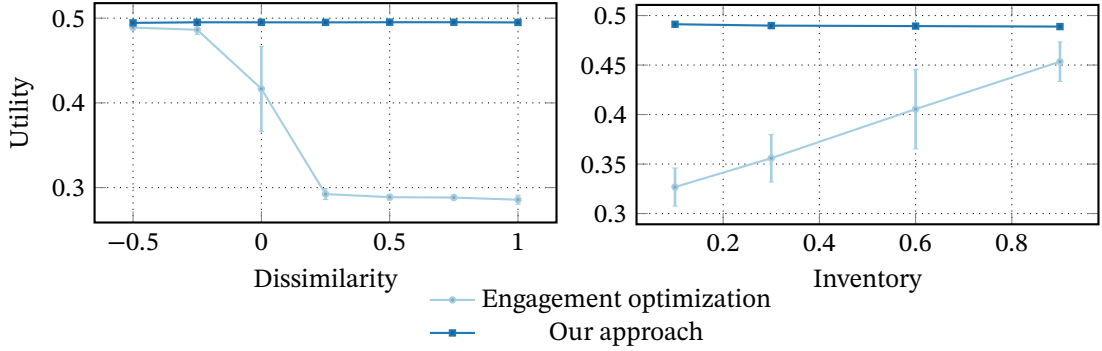
\begin{figure}[t]
    \begin{subfigure}[T]{0.49\linewidth}
        \centering
        \hspace{1em}%
        \begin{tikzpicture}
            \begin{axis}[
                grid=both,
                thick,
                mark size=1pt,
                width=0.95\textwidth,
                height=4.5cm,
                xlabel=Dissimilarity,
                ylabel=Utility,
                legend style={draw=none, at={(1.1, -0.25)}, anchor=north},
                every axis plot post/.append style={error bars/.cd, y dir=both, y explicit, error bar style={line width=1pt}},
            ]
                \addplot table [col sep=comma, x=dissimilarity, y=utility_1, y error=utility_1_std] {./dat/utility_vs_dissimilarity.csv};
                \addplot table [col sep=comma, x=dissimilarity, y=utility_2, y error=utility_2_std] {./dat/utility_vs_dissimilarity.csv};
                \legend{Engagement optimization, Our approach};
            \end{axis}
        \end{tikzpicture}
    \end{subfigure}%
    \begin{subfigure}[T]{0.49\linewidth}
        \hspace{-1.5em}%
        \begin{tikzpicture}
            \begin{axis}[
                grid=both,
                thick,
                mark size=1pt,
                width=.95\textwidth,
                height=4.5cm,
                xlabel=Inventory,
                every axis plot post/.append style={error bars/.cd, y dir=both, y explicit, error bar style={line width=1pt}},
            ]
                \addplot table [col sep=comma, x=inventory, y=utility_1, y error=utility_1_std]{./dat/utility_vs_inventory.csv};
                \addplot table [col sep=comma, x=inventory, y=utility_2, y error=utility_2_std] {./dat/utility_vs_inventory.csv};
            \end{axis}
        \end{tikzpicture}
    \end{subfigure}
    \caption{ The blue curve and the red curve show the session utility obtained by optimizing items with respect to estimated $\e^1+\e^2$ and estimated $\e^2$, respectively, plotted as a function of  $\e^1,\e^2$ dissimilarity on the left and $\m$ inventory on the right.}
    \label{fig:utility}
\end{figure}

\subsection{Effect of gap between $\beta^1$ and $\beta^2$ on estimation error}
In this experiment, our goal is to understand the effect of the difference in $\beta^1$ and $\beta^2$ on the estimation error. Recall that $\beta^1$ and $\beta^2$ represent the decay rate of System-1 and System-2 trigger intensities, respectively. 
We generate item embeddings using the same QR factorization-based procedure described in \cref{sec:error}. 
We set $\beta^2 = 1$ and vary $\beta^2$ in the range $[1,5]$.
We also set the number of samples to be $1024$ (each sample has $1000$ sessions).
We again report the $\ell_2$ norm
of the distance between the estimates and the true values divided by the $\ell_2$ norm of the true values.
Figure~\ref{fig:error} reports the error in estimation 
of model parameters as a function of  $\beta^1-\beta^2$.
One can observe that the error decreases monotonically 
as the gap increases. When $\beta^1 = \beta^2$, the error in estimation of $\u$ and $\m$ is very high.
This is because when moreishness also has a long-term 
effect on System-1 decisions, the two decision-processes look very similar and it is difficult to disentangle them.
Note that the algorithm is still able to estimate $\mu, \beta^1, \beta^2$ reasonably well in this case.
When the gap increases the error reduces sharply  
and converges when $\beta^1 = 4$ and $\beta^2=1$.
This confirms with our intuition that if the decay rates 
for System-1 and System-2 processes are sufficiently different, i.e.,
utilities have a much longer effect on System-2, then we will be able to disentangle the utility from moreishness.

\subsection{Comparing utility and engagement maximization in terms of (dis-)similarity b/w $\e^1$ and $\e^2$}
Here, our goal is to understand whether 
our algorithm is able to maximize  utility 
as compared to engagement optimization.
We  want to compare this as a function of 
the similarity between $\e^1$ and $\e^2$. 
If $\e^1$ and $\e^2$ are well-aligned then one would 
expect that engagement optimization is a good proxy 
for utility maximization. However, if they are not aligned (or even negatively aligned) then one would expect that it is not a good proxy. 
We generate item embeddings in the same manner as Section~\ref{sec:error} using QR factorization of a random matrix.
We generate user embeddings as follows:
we let $\e^2 = Q_1$ and let $\e^1 = -s*Q_1 + Q_2$
where $Q_1$  and $Q_2$ are again the first and second row of $Q$, respectively.
We normalize these vectors to have $\ell_2$ norm $1$.
A positive value of $s$ means that utility and moreishness embeddings are somewhat opposite to each other. This can, for instance, happen when moreishness leads to overuse which leads to lower  utility because of missing out on the outside option  \citep{KleinbergMR22}.
A negative value of $s$ means that utility and moreishness embeddings are somewhat aligned and engagement optimization may perform well in this case.
This can, for instance, happen when high utility items also provide good entertainment to the user.

We then calculate the utility of our approach as well as engagement optimization as a function of the value of $s$.
In order to calculate the utility of our approach 
we find the set $S$ of top $10$ items that have the largest dot product with estimated $\e^2$, and then calculate the 
average utility $\alpha^2 = \phi(\v_S^\top \e^2)$ of the set of items $S$.
We do the same calculation to find the utility for engagement maximization 
except here we maximize the dot product to $\e^1 + \e^2$ \footnote{Note that we do not model engagement explicitly in our work, but our understanding is that both $\u$ and $\m$ combine together in some way to generate the engagement signal. In our experiments we assume that engagement is a function of $\u+\m$, after taking inspiration from the model of \cite{KleinbergMR22}.}, i.e., if we would have recommended using the entangled engagement signal.
Figure~\ref{fig:utility} plots the utility 
as a function of the dissimilarity parameter $s$.
One can observe that our approach achieves the highest possible utility of $0.5$ that is achievable in our setup. This also show that the $\e^2$ embedding is estimated well and using it for content optimization is akin to using the true embedding.
More importantly, it shows that for engagement optimization the utility declines 
sharply when $s$ becomes positive and one will achieve 
almost half the utility even for small misalignment between $\e^1$ and $\e^2$.

\subsection{Comparing utility and engagement maximization in terms of inventory of $\e^2$ items}
In this experiment our goal is to understand the effect of 
availability of items that are aligned with user utility ($\m$).
The idea is that if most of the items available in the inventory are engaging low-utility items then optimizing with respect to $\m$ may yield very different results 
as compared to optimization with respect to $\u+\m$.
On the other hand, if most of the items are aligned with 
user utility then optimizing with respect to $\m$ might give similar results as optimization with respect to $\u+\m$.
To generate the embeddings we again compute the QR 
factorization of a random matrix $A$, i.e. $A = Q \times R$.
We generate user embeddings as: $\e^2 = Q_1$ and $\e^1 = -0.2*Q_1 + Q_2$
where $Q_1$  and $Q_2$ are again the first and second row of $Q$, respectively.
We generate item embeddings as follows:
let $s$ be a parameter ranging in $[0,1]$, 
we draw a random draw from Bernoulli$(s)$ and if it lands as heads we let 
$\v = \e^2 + \epsilon$; otherwise we let $\v = \e^1 + \epsilon$,
where $\epsilon$ is a random noise vector where each dimension is $\mathcal{N}(0,1/10d)$.
Hence, we roughly have $s$ fraction of items that are 
aligned with $\m$ and $1-s$ fraction that are aligned with $\u$.
We normalize all vectors to have $\ell_2$ norm $1$.
We then calculate the utility of our approach as well as engagement optimization as a function of the value of $s$.
In order to calculate the utility of our approach 
and engagement maximization, we follow the same steps 
as the previous section after finding the set of top $10$
items that have the largest dot products with respect 
to the corresponding embeddings. 
Figure~\ref{fig:utility} plots the utility 
as a function of the parameter $s$.
One can observe that our approach achieves the highest possible utility of $0.5$ that is achievable in our setup. 
More importantly, it shows that the utility for engagement optimization approach is very low when the fraction of $\m$-aligned 
items is low.  
Also, the utility for engagement optimization increases monotonically as the fraction of $\m$-aligned items increases which is in-line with our intuition.

%% file: conclusion.tex
\section{Discussion}
\label{sec:discuss}

\paragraph{Alternative Platform Objectives}

In addition to maximizing per-session utility, a platform can also consider the objective of \emph{maximizing average utility over an infinite time-horizon}.
Formally, this objective function is defined as
\[
\lim_{T\rightarrow \infty}
 \frac{1}{T} \sum_{t \in \D} \phi\left( (\m)^\top \v_{S_t}  \right)
 \,,
\]
where $T$ is the time-horizon over which the samples are collected.
An obvious question that comes to mind is whether 
maximizing per-session utility will also result in 
maximization of this objective under our model.
It is easy to see that this is not the case. 
This is because the above objective depends on both the utility per-session as well as  
the number of user sessions within the time-horizon $T$.
Hence, maximizing per-session utility will not be enough if it does not lead to large number of user 
sessions over the time-horizon $T$. For example, if there is an item with good moreishness and good utility, 
then it might be better to recommend this item over an 
item that has the best utility but low moreishness.
This is because the former will maximize the intensity as well as provide good utility
as compared to the latter which provides the best utility but does not provide the good intensity.
Another strategy might be to alternate between high-utility and moreish sessions.  
In this case one needs to consider the optimization/control of this objective across sessions.

One can also consider the objective of maximizing the 
number of daily active users. Under our model this would amount to maximizing the integral of the intensity of the Hawkes process corresponding to each user. 
However, one has to be careful with this objective function as there might also be some undesirable ways to 
maximize it. 
For example, one way to maximize this objective might be to maximize per-session engagement so that the System-1 trigger is high enough to keep bringing the user back to the platform.

\paragraph{Alternative Session Summarizing Techniques}
\label{sec:credit}
In our formulation we summarize each session by taking 
the average of individual item embeddings within the session
and then take the dot product with user embedding to calculate utility/moreishness.
It would also be interesting to consider other ways to summarize a session, for example, by taking a weighted
combination of item embeddings where the weights depend on the engagement signals. 
This will allow us to put more importance to items that 
influence the user more. 
One can also utilize long short term memory (LSTM)
based neural networks for summarizing each session.

\section{Limitations}
\paragraph{Validity of Assumptions on User Behavior}
While our ``dual system'' model has not yet been tested in  
a real-world scenario, it is based on empirical findings about impulsive usage on online platforms \citep{cho2021reflect, lyngs2019self, moser2020impulse} and 
inspired by psychological mechanisms for decision-making 
 \citep{thaler1981economic, akerlof1991procrastination, laibson1997golden, KleinbergMR22}.
As for most modeling scenarios, it is plausible that our assumptions are not satisfied exactly in a real-world application.  
For example, it may be possible that a user keeps returning to the platform even when their long-term goals are not being met.
However, there is empirical evidence suggesting a positive correlation between 
utility and long-term retention \citep{gomez2015netflix, muddiman2019clickbait}.
Hence, we expect that at an aggregate level our method will get better “directional” information about utility than engagement-based methods, even if the model is only approximately correct.
Moreover, as pointed out earlier in this section, our model has added flexibility 
that allows it to be fine-tuned or extended to different settings.

\paragraph{Non-Stationarity of User Arrival Intensity}
In a real-world scenario, it is likely that
the user arrival intensity will change over time.
Firstly, there can be a \emph{seasonality}  
or \emph{time-of-day} effect. Secondly, there can be changes to the recommendation policy which can change the distribution of content on the platform. 
Lastly, the user utility can change over time resulting in changes to the arrival intensity. 
In this work, our focus has been on distinguishing System-1 and System-2 components of user behavior. Hence, we abstracted away from this discussion about non-stationarity.
While some amount of non-stationarity can be handled by our current model, for example, by tuning the frequency of data collection and policy optimization so that observed utilities are approximately stationary,
there can still be scenarios where 
the underlying non-stationarity can confound the inferences made by our model.
However, effects like seasonality are well-explored in temporal point process literature and can be incorporated in our model.
For instance, we can explicitly account for some non-stationarity by adding a time-dependent base intensity $\mu(t)$ to our Hawkes process model.

\paragraph{Known Item Embeddings}
As mentioned in \Cref{sec:problem}, we assume that the item embeddings are known.
This is motivated by the abundant availability of item-level data such as item attributes, audio-visual features and engagement signals.
Similar to matrix factorization techniques,
one can also consider joint learning of item and
user features solely based on the return behavior. However, the joint identifiability and learning becomes more challenging in this case.

\section{Related Work}
The choice inconsistencies exhibited by humans have been 
well-documented and explained in the psychology literature through various mechanisms.
There has been significant work on the dual systems theory 
which posits the existence of two separate decision-processes coexisting with each other
\citep{kahneman2011thinking, smith2000dual, sloman1996empirical, schneider1977controlled, evans2008dual}.
Even though the specific psychology mechanisms gets nuanced with several additional connotations attached to System-1 and System-2,
we rely on an abstraction where 
System-1 is the myopic decision-maker and  
System-2 optimizes for long-term goals.
There is also other work in economics and computer science that considers 
issues with time inconsistency and self-control  in 
human decision-making \citep{akerlof1991procrastination, thaler1981economic, laibson1997golden, o1999doing, kleinberg2014time, lattimore2014general}. 
The phenomenon of choice inconsistency and lack of self-control has also been documented empirically in various settings
\citep{milkman2010ll, cryder2017charity, milkman2009highbrow, gruning2023directing}.

The literature on recommendation systems has received a lot of recent attention towards misalignment between engagement and utility. \cite{MilliBH21} consider the goal of scoring different engagement signals such as like, share, watch time, etc, in terms of their correlation with utility.  
However, their work ultimately uses engagement signals to measure utility, whereas we use longer-term return probabilities.
\cite{milli2023choosing} also consider weighting 
different engagement signals from the perspective of  alignment with utility, strategy-robustness and ease of estimation.
\cite{KleinbergMR22} propose a model 
of user interaction within a session and illustrate the 
pitfalls of engagement optimization when users make decisions according to both System-1 and System-2. 
The main difference between our work and theirs is that we model the decision of users to start a new session, whereas \cite{KleinbergMR22} are mainly concerned with the decisions to continue a session once it is already started.
The HCI literature has also explored the question of 
understanding user utility beyond engagement optimization. Various methods have been suggested such as eliciting explicit or implicit feedback from users about their
experience on the platform \citep{LyngsBKS18, LyngsLSBSIKS19}.
There has also been work on value-alignment in recommender systems \citep{stray2022building, Stray+21} where the goal  is to optimize for 
different values such as  diversity, fairness, safety, etc.,
in addition to engagement.

Over the last several years there has also been a focus on optimizing long-term objectives in recommendation systems. 
There has been work on optimizing short-term objectives under long-term constraints such as fairness, diversity, legal requirements  \citep{brantley2023ranking, UsunierDD22, CelisKSV19, MorikSHJ20}.
However, these long-term constraints are explicitly specified by the platform or policy requirement instead of being implicitly specified by the user.
There is also been work in the multi-armed bandits 
literature that considers optimizing for long-term rewards in the context of recommendation systems \citep{WuWHS17, McDonald+23}. 
Finally, the reinforcement learning (RL) literature has 
also devoted significant attention towards maximizing long-term reward metrics in recommendation systems \citep{ZouXDS0Y19, ZhaoXTY19}. 
These works, however, consider explicit optimization of (appropriately defined) long-term reward, whereas we use 
long-term return probabilities of users as a mere proxy for true user utility.
Moreover, these works do not consider the 
choice inconsistencies in behavior exhibited by the users
and assume that their actions are in accordance with their utility.
On the other hand we differentiate between utility-driven and impulse-driven behaviors with the 
goal of optimizing content with respect to utility.

Temporal point-processes are a fundamental tool for spatial 
data analysis and have found application in a  
wide-range of domains such as finance, epidemiology, seismology,  and computational neuroscience \citep{daley2003introduction}.
Recently, they have also been studied in the context of recommendation systems. \cite{WangDTS16} studied the co-evolution dynamics of user and item embeddings through the lens Hawkes processes.  The most closely related to our work 
is \cite{jing2017neural} who model the return probabilities of users based on a LSTM-based point process.
However, apart from differences in specific modeling choices, the main difference is that \cite{jing2017neural} assume that all choices made by the users are in accordance with their utility, 
whereas we differentiate between utility-driven and impulse-driven behaviors.
There has also been substantial work in modeling the activities of  
users on social media using point processes, e.g., see \cite{Gomez-RodriguezBS11,nickel2021modeling}.
We refer the reader to a tutorial by \cite{rodriguez2018learning} and surveys by \cite{yan2019recent} and \cite{shchur2021neural}
for other machine learning applications.

\section{Conclusion}
\label{sec:conc}
In this paper we explore a new approach to recommender systems that does not optimize for content using engagement signals. This is because of the risk of optimizing for impulsive (System-1) behavior when using  
engagement signals. 
Instead, our focus is on using long-term arrival rates as a way to understand 
the utility of content for a user.  
We design a generative model for user arrival rates based on a self-exciting Hawkes process where both System-1 and System-2 together govern the arrival rates.
Positive utility in the current session has a lasting effect on future System-2 arrival rate,
while moreishness only effects the System-1 arrival rates in the near future.
Using samples from this process allows us to disentangle the effects of System-1 behavior and System-2 behavior
and allows us to optimize content with respect to 
utility.
Using experiments on synthetic data we show that 
the utility obtained using our approach is much 
higher than the utility obtained using engagement optimization.

We believe that our paper can provide important insights into  utility maximization in recommendation systems and can lead to more work in this area. An exciting direction for future work is to look at 
other signals in addition to user arrival rates 
and understand if there is a way to combine these signals 
with engagement signals which are more abundantly available.
It would also be interesting to look at other ways of 
summarizing a session as compared to taking a simple average.
It would be interesting to strengthen our theoretical 
results by providing an analysis for the case where 
the length of the session is correlated with 
System-2 utility, and hence, there is an entanglement between the observed session lengths and the observed arrival times.

%% file: appendix.tex
\section{Additional Lemmas on Identifiability and Consistency}
\label{app:technical}
The following lemma establishes the identifiability of a simple trigger function $\kappa(t)$.

\begin{lemma}
\label{lem:trigger_1}
The triggering function 
$$\kappa(t) = \beta^1\alpha^1\exp(-\beta^1 t) + \beta^2 \alpha^2 \exp(-\beta^2 t)$$ defined over domain $\R_+$ is identifiable if $\beta^1\neq \beta^2$ and $\alpha^1, \alpha^2, \beta, \beta^2 > 0$.
\end{lemma}
\begin{proof}
Consider the following density function 
$$f(t) = \frac{\alpha^1}{\alpha^1+ \alpha^2}\beta^1 \exp(-\beta^1t) +\frac{\alpha^2}{\alpha^1+ \alpha^2} \beta^2 \exp(-\beta^2 t)$$
for $t \geq 0$ and $0$ otherwise.
It is easy to verify that this is the probability density function of a mixture of exponential distributions. 
The classic result of \cite{Teicher61} shows that 
mixtures of exponential distributions are 
identifiable. Using this result, and the fact that the 
mapping from $\alpha^1$ to $\alpha^1/(\alpha^1+\alpha^2)$ 
is one-to-one (given fixed $\alpha^2$), implies that the triggering function $\kappa(t)$ is identifiable. 
\end{proof}